\documentclass[
preprint,
showpacs,
showkeys
]{revtex4-1}
\usepackage{amsthm}
\theoremstyle{plain}
\newtheorem*{theorem*}{Theorem}

\newtheorem*{definition*}{Definition}

\newtheorem*{lemma*}{Lemma}

\usepackage{braket}
\usepackage{delarray}
\usepackage{here}

\usepackage{txfonts}

\usepackage[dvipdfmx]{graphicx}
\usepackage{bm}
\usepackage{color}
\usepackage{ulem}

\newcommand{\be}{\begin{eqnarray}}
\newcommand{\ee}{\end{eqnarray}}
\newcommand{\ba}{\begin{array}}
\newcommand{\ea}{\end{array}}
\newcommand{\bmat}{\left(\begin{array}}
\newcommand{\emat}{\end{array}\right)}

\begin{document}
\title{
Analysis on the von Neumann entropy under the measurement-based feedback control}

\author{Kohei Kobayashi$^1$}

\affiliation{
$^1$Global Research Center for  Quantum Information Science, National Institute of Informatics,
 2-1-2 Hitotsubashi,  Chiyoda-ku, Tokyo 101-8340, Japan}

\begin{abstract}

The measurement-based feedback (MBF) control  offers several powerful means for preparing 
the desired target quantum state.
Therefore, it is important to investigate fundamental properties of MBF.
In particular, how the entropy of the controlled system under the MBF behaves is of great interest.
In this study, we examine this problem
by deriving a sufficient condition that the time derivative of the von Neumann entropy 
is nonnegative under the MBF control.
This result is rigorously characterized by the variance of the system observable and the quantumness of a given decoherence.
We show the validity of the result and physical interpretation in the example of qubit stabilizing.


\end{abstract}
\date{\today}
\maketitle

\section{Introduction}

Due to the requirements of rapidly development of quantum technologies, 
 the control of quantum systems has become one of the important topics in engineering. 
 Theoretical advances in quantum control play an essential role in realizing quantum information technologies, e.g., 
quantum computation, quantum communication, quantum teleportation, and quantum metrology, which 
achieve high performances beyond the limits of existing classical technologies \cite{Nielsen}.

In particular, quantum feedback control offers several powerful means for preparing the desired target state.
The measurement-based feedback (MBF) control is a well established methodology that can be used for preparing and protecting 
a desired quantum state.
The basic strategy is to control the dynamics of the system by using the information obtained by the measurement;
that is, we first measure the target system and fed the measurement result back to control it.
The evolution of the system is described by the stochastic differential equation, called stochastic master equation (SME).
The theory of MBF has been well formulated \cite{Stockton, Handel, Geremia, Yanagisawa, Hara, Mirrahimi, Nurdin} and
some notable experiments have been demonstrated in recent years \cite{Vijay, Gourgy, Cox}.

On the other hand, it is also important to investigate fundamental properties of the MBF in general settings.
In the literatures, several studies have been reported;
the advantages of MBF compared with the other control method have been investigated \cite{Yamamoto},
the controllability of the MBF in the presence of decoherence was characterized \cite{Bo, Kobayashi},
the lower bound of the time required to stabilize the target state by the MBF was given \cite{Pedro}, and so on.
In particular, it is of great interest to see how the entropy of the conditioned quantum system by the MBF control.
Therefore, we have a natural question; 
how does the entropy in the controlled quantum system by the MBF behave under the decoherence?

In this study, we partially give an answer to this question; 
more precisely, we derive a sufficient condition that the time derivative of the von Neumann entropy 
is nonnegative under the MBF control.
This result is rigorously characterized by the variance of the observable of the system and the quantumness of a given decoherence.
Unfortunately, it is noted that the result is sufficient condition not necessary one.
Through a simple but illustrative example of the qubit system, 
we show the validity of the result and physical interpretation.

\section{Entropy bound}

\subsection{Controlled quantum dynamics}

Let us explain a time evolution of the Markovian open quantum system in a typical setting of MBF.
The state of a quantum system is represented by the density matrix $\rho_t$, which satisfies 
 $\rho^\dagger=\rho$, $\rho\geq 0$, and ${\rm Tr}[\rho]=1$.
When the system is conditioned by the continous measurement, 
the general dynamics of $\rho_t$ is represented as follows \cite{Nurdin}:

\begin{equation}
\label{SME}
d \rho_t = -i[u_t H , \rho_t]dt + \mathcal{D}[L ]\rho_tdt + \mathcal{D}[ M ]\rho_tdt 
+ \mathcal{H}[ L] \rho_tdW_t.
\end{equation}
This is known as the stochastic master equation (SME).
Here the superoperators $\mathcal{D}$ and $\mathcal{H}$ are defined by
 \begin{eqnarray}
 \mathcal{D}[A] \rho &=& A\rho A^\dagger - \frac{1}{2}A^\dagger A\rho-  \frac{1}{2}\rho A^\dagger A, \\
 \mathcal{H}[A] \rho &=& A\rho +\rho A^\dagger -{\rm Tr}[(A +A^\dagger )\rho]\rho.
\end{eqnarray}
 $H$ is the system Hamiltonian, and
 $L$ and $M$ are the Lindblad operators describing the interaction between the system and the environment. 
$L$  represents the controllable coupling; for instance, in the MBF setting, $L$ corresponds to the probe for measurement.
On the other hand,  $M$ represents the undesirable coupling induced by the decoherence.
(we set $\hbar=1$ in the following).
$dW_t=dy_t-{\rm Tr}[(L+L^\dagger )\rho_t]dt$ is the infinitesimal
 Wiener increment representing the innovation process based on the measurement result $y_t$.
The goal of the MBF is to determine the control input $u_t$ as a function of $\rho_t$, to achieve a certain target.
If $M=0$ and $L=L^\dagger$, Eq. (\ref{SME}) represents the ideal dynamics that steer the state to an arbitrary eigenstate of $L$.

Here let us focus on the unconditional state $\mathbb{E}[ \rho_t]$, which is the ensemble average of $\rho_t$
over all the measurement results.
Due to the fact that $\mathbb{E}[ dW_t]=0$ and  the control sequence $u$ is a function of the state $\rho_t$,
Eq. (\ref{SME}) becomes the following master equation:

\begin{equation}
\label{ME}
\frac{d\mathbb{E}[ \rho_t] }{dt} = -i[ H , \mathbb{E}[u(\rho_t) \rho_t] ] 
+ \mathcal{D}[L ] \mathbb{E}[ \rho_t] + \mathcal{D}[M] \mathbb{E}[ \rho_t].
\end{equation}

Note that Eq. (\ref{me2}) is not linear with respect to $\langle \rho_t\rangle$.
If $u$ is independent of  $\rho_t$, Eq. (\ref{ME2}) reduces to the linear equation:

\begin{equation}
\label{me3}
\frac{d \mathbb{E}[ \rho_t]  }{dt} = -i[ u_t H, \mathbb{E}[ \rho_t]] 
+ \mathcal{D}[L] \mathbb{E}[ \rho_t] + \mathcal{D}[ M] \mathbb{E}[ \rho_t].
\end{equation}

\subsection{Main result}

We introduce the von Neumann entropy to measure the degree of decoherence affecting on the process
of controlling the system state to a target state. 
The von Neumann entropy in $\rho_t$ can be defined by

\begin{equation}
\label{vne}
S_t =-{\rm Tr}\left[ \rho_t \ln \rho_t\right], \ \ \ 0\leq S_t \leq \ln d,
\end{equation}
where $d$ is the rank of the quantum system.
The von Neumann entropy quantifies the amount of information contained 
in a state.
$S_t$ takes zero when $\rho_t$ is pure, 
and the maximum is achieved only when the system is maximally mixed $\rho_t=I/d$.  

Remarkably, the von Neumann entropy is a quantum analogue of the Shanonn entropy \cite{Shannon}:
\begin{equation}
S_t =-\sum_j \lambda_{j, t} \ln \lambda_{j, t},
\end{equation}
where $\lambda_j$ are the time-dependent eigenvalue of $\rho_t$.

Let us consider the evolution of the von Neumann entropy.
The infinitesimal change of the von Neumann entropy is given by 

\begin{eqnarray}
\label{dS}
dS_t &=& 
-{\rm Tr}[d \rho_t \ln \rho_t ] - {\rm Tr}[ \rho^{-1}_t (d \rho_t)^2],     \nonumber \\
&=& -{\rm Tr}\left\{ \mathcal{D}[ L] \rho_t \ln \rho_t \right\}dt 
- {\rm Tr} \left\{ \mathcal{D}[M] \rho_t \ln \rho_t \right\} dt 
- {\rm Tr}\left\{ \rho^{-1}_t (\mathcal{H}[ L]  \rho_t )^2 \right\} dt \nonumber \\
&\ \ \ \ & - {\rm Tr}\left\{  \mathcal{H}[ L] \rho_t \ln \rho_t \right\} dW_t, 
\end{eqnarray}

where we have used 

\begin{eqnarray}
&& {\rm Tr}[d \rho_t]=0,  \\ 
&&  {\rm Tr}\left( [H, \rho_t]\ln \rho_t \right)=0.
\end{eqnarray}

Using the relation $[\rho_t, \ln \rho_t]=0$, the first term of (\ref{dS}) becomes

\begin{eqnarray}
\label{derive1}
-{\rm Tr}\left\{ \mathcal{D}[L] \rho_t \ln \rho_t \right\}  
&=& {\rm Tr}\left\{ (L^2 \rho_t - L \rho_t L)\ln \rho_t  \right\}.
\end{eqnarray}

According to the result derived in \cite{Abe}, Eq. (\ref{derive1}) has the lower bound:

\begin{eqnarray}
\label{derive2}
-{\rm Tr}\left\{ \mathcal{D}[L] \rho_t \ln \rho_t\right\}dt 
\geq  {\rm Tr}\left\{ [L^\dagger, L] \rho_t \right\} = \langle [L^\dagger, L]\rangle = 0.
\end{eqnarray}

Likewise, the second term of (\ref{dS}) also has the lower bound:
\begin{eqnarray}
\label{derive3}
-{\rm Tr}\left\{ \mathcal{D}[ M]\rho_t \ln \rho_t\right\} 
 \geq \langle [M^\dagger, M]\rangle.
\end{eqnarray}

To calculate the bound of the third term of (\ref{dS}), 
we consider 

\begin{eqnarray}
\left(\mathcal{H}[L]\rho_t \right)^2
&=& \left( L \rho_t +\rho_t L-2{\rm Tr}[L\rho_t]\rho_t \right)^2  \nonumber   \\
&=&L \rho_tL\rho_t +  L\rho_t^2 L + \rho_t L^2 \rho_t
+ \rho_t  L \rho_t L
-2{\rm Tr}[L \rho_t](\rho_t L\rho_t +\rho_t^2 L )  \nonumber   \\
&\ \ \ \ &-2{\rm Tr}[L \rho_t](\rho_t L \rho_t + L \rho_t^2 )   
+4{\rm Tr}[ L \rho_t]^2.
\end{eqnarray}

Therefore, the third term is given as follows:

\begin{eqnarray}
\label{derive4}
 {\rm Tr}\left\{ \rho^{-1}_t (\mathcal{H}[L]  \rho_t )^2\right\}  
&=& 4{\rm Tr}[L^2 \rho_t] -4{\rm Tr}[ L\rho_t ]^2  \nonumber   \\
&=&4{\rm Var}[\rho_t],
\end{eqnarray}
where we defined the variance of $L$ in the quantum system $\rho_t$ as ${\rm Var}[\rho_t]$. 

Combining (\ref{derive2}), (\ref{derive3}), (\ref{derive4}) with (\ref{dS})
 and taking the expectation with respect to $W_t$, we have the following inequality:

\begin{eqnarray}
\label{dSdt}
\frac{d \mathbb{E}[ S_t] }{dt} &\geq& 
 {\rm Tr}\left( [M^\dagger, M] \mathbb{E}[ \rho_t] \right) 
-4{\rm Tr}\left(L^2 \mathbb{E}[ \rho_t] \right) + 4 \mathbb{E}\left[ {\rm Tr}[L \rho_t]^2 \right]  \nonumber   \\
&\geq & \mathbb{E}\left[\langle  [M^\dagger, M] \rangle \right] 
 -4  {\rm Tr}\left( L^2 \mathbb{E}[ \rho_t]  \right)
+4 \mathbb{E}\left[  {\rm Tr}[ L\rho_t ] \right]^2   \nonumber   \\
&=& \mathbb{E}[\langle  [M^\dagger, M] \rangle]  -4{\rm Var}\left[ \mathbb{E}[\rho_t] \right],
\end{eqnarray}
where we used the inequality $\mathbb{E}[ x^2] \geq \mathbb{E}[x]^2$.

We find that 
the evolution of the von Neumann entropy is characterized by 
the quantumness of the decoherence operator and the variance of the observable in the averaging state.
Therefore, we find that the effect of the continuous measurement prevented the von Neumann entropy  from monotonically increasing.

First, we assume that $M^\dagger =M$, i.e., $M$ represent the observable.
Then we have
\begin{eqnarray}
\frac{d \mathbb{E}[ S_t] }{dt} \geq -4{\rm Var}\left[ \mathbb{E}[\rho_t] \right].
\end{eqnarray}
Thus, it is not clear whether the time derivative of the entropy is positive or negative.
This is attributed that it is impossible to distinct $M$ from the measurement operator $L$ in the expression.
Furthermore, if $L=0$, we find that $d \mathbb{E}[ S_t] /dt \geq 0$.
In this case, the system converges to the one of the eigenstate of $M$.

On the other hand, when $M^\dagger \neq M$,
the righthand side of  (\ref{dSdt}) is greater than zero if
\begin{eqnarray}
\label{suff}
\mathbb{E}[\langle  [M^\dagger, M] \rangle] \geq  4{\rm Var} \left[ \mathbb{E}[\rho_t] \right].
\end{eqnarray}
As a result, the system operators $L$, $M$, and $\mathbb{E}[\rho_t] $ satisfy this relation, 
it is guaranteed that the von Neumann entropy is monotonically increasing or conserved.
Importantly, we would like to point out that Eq. (\ref{suff}) is the sufficient condition not the necessary one.

\section{Example: Qubit stabilization}

Let us apply the discussion developed above to a simple but illustrative example of 
a qubit such as a two-level system consisting of the excited state 
$|0\rangle=(1, 0)^\top$ and the ground state $|1\rangle=(0, 1)^\top$.
In this case, the density matrix is parametrized as

\begin{eqnarray}
\rho_t =\frac{1}{2}\left(x_t \sigma_x +y_t \sigma_y +z_t \sigma_z   \right),
\end{eqnarray} 
where $x_t$, $y_t$, and $z_t$ are time-dependent real scalars.
Here positive semidefiniteness of $\rho_t$ leads to the condition $x^2_t+y^2_t+z^2_t \leq 1$.
$\sigma_x =|1\rangle \langle 0|+ |0 \rangle \langle 1|$, 
$\sigma_y =-i|1\rangle \langle 0|+ i|0 \rangle \langle 1|$, and
 $\sigma_z =|0\rangle \langle 0|- |1 \rangle \langle 1|$ are the Pauli matrices.

Let the initial state and the ideal target final state be $\rho_0=|+\rangle\langle +|$, and 
 $\rho_T=| 0\rangle\langle 0|$, where $|+\rangle=(|0\rangle + |1\rangle)/\sqrt{2}$ is the superposition of $|0\rangle$ and $|1\rangle$.
 
Here we consider the operators:
\begin{eqnarray}
H=\sigma_y, \ \ \ L=\sqrt{\kappa}\sigma_z, \ \ \ M=\sqrt{\gamma}\sigma_-,
\end{eqnarray} 
where $\sigma_-=|1\rangle\langle0 |$ and $\sigma_+=|0\rangle\langle1 |$ 
are the lowering and raising  matrices.
This is a typical MBF control setting \cite{Handel, Vijay, Gourgy}.
$L$ represents the dispersive coupling between the qubit and the probe, which enables us 
to monitor the system by measuring the probe output and 
perform a feedback control through the Hamiltonian $u_tH$.
$M$ represents the energy decay from $|0\rangle$ to $|1\rangle$, 
which is interpreted as a spontaneous emission for a two-level system.

In this setup, we obtain the following:
\begin{eqnarray}
 \mathbb{E}[ \langle [M^\dagger, M]\rangle]  
 &=& \gamma{\rm Tr}\left( [\sigma_+, \sigma_-] \mathbb{E}[\rho_t] \right)   \nonumber \\
&=& \gamma \mathbb{E}[z_t], \\
 {\rm Var} \left[ \mathbb{E}[\rho_t] \right] 
&=& \kappa{\rm Tr} (\sigma^2_z \mathbb{E}[\rho_t] ) -\kappa{\rm Tr}( \sigma_z  \mathbb{E}[\rho_t] )^2  \nonumber \\
&=& \kappa-\kappa \mathbb{E}[z_t] ^2.
\end{eqnarray} 
To make the analysis easier, we set $\gamma=\alpha\kappa$ with the positive constant $\alpha>0$. 
Then, from the expression (\ref{suff}),  we have the inequality
 \begin{eqnarray}
 \label{result}
 \frac{-\alpha +\sqrt{64+\alpha^2}}{8}  \leq \mathbb{E}[z_t] \leq 1,
\end{eqnarray} 
for $d \mathbb{E}[S_t]/dt\geq0$.
This result gives a simple physical interpretation.
If $\alpha=0$ (i.e., no decoherence), the lower bound of Eq. (\ref{result}) takes $1$, 
and thus, it is unclear whether the von Neumann entropy is monotonically increasing or not, for all $\mathbb{E}[z_t]\in[-1, 1]$. 
Meanwhile, in the limit $\alpha\to\infty$, this lower bound approximates zero.
This result seems that the entropy increases easily when the decoherence is sufficiently large, 
but at the same time, it becomes difficult for $\mathbb{E}[z_t]$ to satisfy this range.


\section{Conclusion}

We investigated the von Neumann entropy in the controlled quantum system by the MBF under the decoherence.
We rigorously showed that the time derivative of the von Neumann entropy is estimated from below by the variance 
of the observable that is continously measured and the quantumness of decoherence.
From the obtained expression, we derived the sufficient condition characterizing a behaviour of the von Neumann entropy and
analytically examined its validity through the qubit stabilization.
The discussion developed in this paper is expected to be useful for studying the dynamics of quantum properties in the presence of decoherence.
A remaining work is to explore a fundamental limit on the von Neumann entropy under the MBF control with decoherence.

This work was supported by MEXT Quantum Leap Flagship Program Grant JPMXS0120351339.

\appendix

\section{Proof of Eq. (6)}

To show the inequality Eq. (6), we use the diagonalization of the density matrix
\begin{equation}
\rho =U\Lambda U^\dagger,
\end{equation}
where $U$ is an appropriate unitary matrix and $\Lambda ={\rm diag}\{\lambda_{1},\cdots, \lambda_{d} \}$ with the eigenvalues
$1\geq \lambda_{1} \geq \cdots \geq \lambda_{d} \geq 0$.
Then,

\begin{eqnarray}
-\ln\rho &=& -U \ln \Lambda U^\dagger  \nonumber \\
&=& U {\rm diag}\{ -\ln\lambda_{1},\cdots,  -\ln \lambda_{d} \} U^\dagger  \nonumber \\
&\geq & U {\rm diag}\{ 1-\lambda_{1},\cdots,  1-\lambda_{d} \} U^\dagger  \nonumber \\
&=& U\left(I-\Lambda \right)U^\dagger=I-\rho,
\end{eqnarray}
where we used $-\ln x \geq 1-x$ for $x>0$.


\end{document}